\documentclass{modified_aipproc}
\usepackage{amssymb}
\layoutstyle{8x11double}

\newcommand{\be}{\begin{equation}}
\newcommand{\ee}{\end{equation}}
\newcommand{\bea}{\begin{eqnarray}}
\newcommand{\eea}{\end{eqnarray}}
\newcommand{\nn}{\nonumber}
\newcommand{\lsim}{
\mathrel{\hbox{\rlap{\hbox{\lower4pt\hbox{$\sim$}}}\hbox{$<$}}}}
\newcommand{\gsim}{
\mathrel{\hbox{\rlap{\hbox{\lower4pt\hbox{$\sim$}}}\hbox{$>$}}}}

\begin{document}
\title{Properties of the Lightest Neutralino in MSSM Extensions
\footnote{Proceedings for PASCOS 2005 (Gyeongju, Korea) talk given by
H.S. Lee. For a full paper by the authors, see Ref. \cite{main}}}
\author{Vernon Barger}{address={Department of Physics, University of Wisconsin,
Madison, WI 53706}}
\author{Paul Langacker}{address={Department of Physics and Astronomy,
University of Pennsylvania, Philadelphia, PA 19104}}
\author{Hye-Sung Lee}{address={Department of Physics, University of Wisconsin,
Madison, WI 53706},altaddress={Department of Physics, University of
Florida, Gainesville, FL 32611}}

\begin{abstract}
\noindent We study neutralino sectors in extensions of the MSSM that
dynamically generate the $\mu$-term. The extra neutralino states are
superpartners of the Higgs singlets and/or additional gauge bosons.
The extended models may have distinct lightest neutralino properties
which can have important influences on their phenomenology. We
consider constraints on the lightest neutralino from LEP, Tevatron,
and $(g-2)_\mu$ measurements and the relic density of the dark
matter. The lightest neutralino can be extremely light and/or
dominated by its singlino component, which does not couple directly
to SM particles except Higgs doublets.
\end{abstract}
\maketitle

\section{Introduction}
\label{introduction}
Extensive studies of the Minimal version of the Supersymmetric SM
(MSSM) show that its lightest neutralino has the right ranges of
mass and interaction strength to be a good cold dark matter (CDM)
candidate \cite{MSSMCDM}. Although the MSSM may be the optimal low
energy Supersymmetric model with minimal extension of the fields and
symmetry, the model may not fully describe the TeV scale physics. It
is important to see if other versions of the Supersymmetric SM can
also give acceptable CDM.

We consider  various extended MSSM models that resolve the
$\mu$-problem \cite{muproblem} and compare their lightest neutralino
properties to that of the MSSM. In these beyond-MSSM models, at
least one Higgs singlet is commonly present and generates an
effective $\mu$ parameter when the associated symmetry is broken at
the EW or TeV scale. Because of the superpartners of the Higgs
singlets or the extra gauge boson, the neutralino sector in these
models may be significantly different from that of the MSSM. We
investigate the mass and the coupling of the lightest neutralino of
each model allowed by the model parameters and the experimental
data.

\section{Models}
\label{models}
\begin{table}[t]
\caption{Higgses and Neutralinos of the MSSM and its extensions
\label{table:model}}
\begin{tabular}{c|c|l|l|l}
\hline
Model   & Symmetry & Superpotential & ~Higgses (CP even, CP odd, charged) & ~Neutralinos \\
\hline
MSSM    & --                              & $\mu H_1 H_2$ & $H_1^0, H_2^0, A^0, H^\pm$ & $\tilde B, \tilde W_3, \tilde H_1^0, \tilde H_2^0$ \\
NMSSM   & $\mathbb Z_3$                   & $h_s S H_1 H_2 + \frac{\kappa}{3} S^3 $   & $+~ H_3^0, A_2^0$           & $+~ \tilde S$ \\
nMSSM   & $\mathbb Z^R_5, \mathbb Z^R_7$  & $h_s S H_1 H_2 + \alpha S$                & $+~ H_3^0, A_2^0$           & $+~ \tilde S$ \\
UMSSM   & $U(1)'$                         & $h_s S H_1 H_2$                           & $+~ H_3^0$                  & $+~ \tilde S, \tilde Z'$ \\
S-model & $U(1)'$                         & $h_s S H_1 H_2 + \lambda_s S_1 S_2 S_3$ & $+~ H_3^0, H_4^0, H_5^0, H_6^0, A_2^0, A_3^0, A_4^0$ & $+~ \tilde S, \tilde Z', \tilde S_1, \tilde S_2, \tilde S_3$ \\
\hline
\end{tabular}
\end{table}

The extended MSSM models that we consider are the Next-to-Minimal
Supersymmetric SM (NMSSM) \cite{NMSSM}, the Minimal Non-minimal
Supersymmetric SM (MNSSM) a.k.a. the nearly Minimal Supersymmetric
SM (nMSSM) \cite{nMSSM}, the $U(1)'$-extended Minimal Supersymmetric
SM (UMSSM) \cite{UMSSM}, and the $U(1)'$-extended Supersymmetric SM
with a secluded $U(1)'$-breaking sector (S-model) \cite{S-model}.
All of these extended models prevent the $\mu$-term ($\mu H_1 H_2$)
and allow an effective $\mu$-term ($S H_1 H_2$) through a vacuum
expectation value (VEV) $\left< S \right>$ of a Higgs singlet
associated with a new symmetry.

The NMSSM assumes a discrete symmetry ${\mathbb Z_3}$ to avoid the
$\mu$-term, but allows an $S^3$ term in the superpotential. The NMSSM
is one of the simplest extensions of the MSSM, but the ${\mathbb
Z_3}$ symmetry predicts domain walls which are not observed
\cite{domainwall}.
The nMSSM was devised to avoid the domain wall problem while
maintaining a discrete symmetry. The term $\alpha S$ in the nMSSM is
a loop-generated tadpole term that breaks the discrete symmetry.
The UMSSM uses an Abelian gauge symmetry instead of a discrete
symmetry and thus is free from the domain wall problem. The new
gauge symmetry introduces a new gauge boson $Z'$.
The S-model was introduced to resolve tension between the EW scale
$\mu_{\rm eff}$ and the heavy $Z'$ (up to multi-TeV scale). It is
basically the extension of the UMSSM with 3 additional Higgs
singlets to provide additional contributions to the $Z'$ mass while
keeping $\mu_{\rm eff} = h_s \left< S \right>$ at the EW scale.

The models are listed in Table \ref{table:model} with adopted
symmetries, superpotentials of Higgses, and the Higgses
and neutralinos. The other parts of the superpotentials are the
Yukawa terms of the MSSM and possible extra terms related to the
exotic chiral fields in the $U(1)'$ models needed to cancel
anomalies. In the $U(1)'$-extended model, the addition of one Higgs
singlet does not give an additional CP odd Higgs since a goldstone
boson is absorbed to be the longitudinal mode of the massive $U(1)'$
gauge boson, $Z'$.

For easy comparisons, we use the common notation of $h_s$ for the coefficient of the $S
H_1 H_2$ term in each model. In every model the
dynamically generated effective $\mu$ parameter is given by \bea
\mu_{\rm eff} = h_s \left< S \right> \eea and therefore the VEV of
the Higgs singlet or the symmetry breaking scale needs to be at the
EW/TeV scale.

\section{Neutralino mass matrices}
With the superpartners of the Higgs singlet ($S$) and the $U(1)'$
gauge boson ($Z'$), the UMSSM has 6 neutralinos (MSSM components +
$\tilde S$ + $\tilde Z'$) and its mass matrix ($M_{\chi^0}$) in the
basis of $\{\tilde{B}$, $\tilde{W}_3$, $\tilde{H}_1^0$,
$\tilde{H}_2^0$, $\tilde{S}$, $\tilde{Z'} \}$ is given by \bea
\left( \matrix{ M_1 & 0 & - \frac{g_1 v_1}{2} & \frac{g_1 v_2}{ 2} &
0 & 0 \cr 0 & M_2 & \frac{g_2 v_1}{2} & - \frac{g_2 v_2}{2} & 0 & 0
\cr - \frac{g_1 v_1}{2} & \frac{g_2 v_1}{2} & 0 & - \mu_{\rm eff} &
- \frac{\mu_{\rm eff} v_2}{s} & \xi_{H_1} v_1 \cr \frac{g_1 v_2}{2}
& - \frac{g_2 v_2}{2} & - \mu_{\rm eff} & 0 & - \frac{\mu_{\rm eff}
v_1}{s} & \xi_{H_2} v_2 \cr 0 & 0 & - \frac{\mu_{\rm eff} v_2}{s} &
- \frac{\mu_{\rm eff} v_1}{s} & 0 & \xi_S s \cr 0 & 0 & \xi_{H_1}
v_1 & \xi_{H_2} v_2 & \xi_S s & M_{1'} } \right). \nn \eea where
$\xi_\phi \equiv g_{Z'} Q'_\phi$, $\langle S
\rangle \equiv \frac{s}{\sqrt{2}}$, and $\langle H_i^0 \rangle
\equiv \frac{v_i}{\sqrt{2}}$ with $\sqrt{v_1^2 + v_2^2} \equiv v
\simeq 246$ GeV. The gauge couplings are $g_1 = e / \cos\theta_W$
and $g_2 = e / \sin\theta_W$. For the $U(1)'$ gauge coupling
constant $g_{Z'}$ and the $U(1)'$ charge $Q'$, we use the Grand
Unification Theory (GUT) motivated gauge coupling and the
$\eta$-model charge assignments in our numerical analysis
\cite{main}. The $U(1)'$ charges should satisfy \be Q'_{H_1} +
Q'_{H_2} \neq 0 \qquad Q'_{H_1} + Q'_{H_2} + Q'_S = 0 \ee in order
to replace the $\mu$-term with the effective $\mu$-term dynamically
generated by the Higgs singlet $S$.

The first $5 \times 5$ submatrix corresponds to the nMSSM limit (or
NMSSM limit in the special case $\kappa = 0$) that can be obtained
by taking $M_{1'} \gg {\cal O}$(EW). With a $(5,5)$ entry of $\sqrt{2}
\kappa s$ from $S^3$ term in the superpotential, this would be the
NMSSM limit.

The first $4 \times 4$ submatrix corresponds to the MSSM limit,
which can be obtained by taking $s \gg \cal{O}$(EW) on top of the
nMSSM limit.

The S-model has 9 neutralinos (UMSSM components + $\tilde S_1$,
$\tilde S_2$, $\tilde S_3$), and its mass matrix has 3 more
columns/rows added to UMSSM neutralino mass matrix. The UMSSM limit
can be realized by taking $s_{1, 2, 3} \gg {\cal O}$(EW) with
$\lambda_s$ comparable to gauge couplings. However, the most
realistic case is for small $\lambda_s$ and large $s_{1, 2, 3}$
\cite{S-model}, in which case four of the neutralinos, consisting
almost entirely of $\tilde Z'$, $\tilde S_1$, $\tilde S_2$, and
$\tilde S_3$, essentially decouple from the other states. Since the full
$9 \times 9$ matrix has a number of free parameters, we consider
only this decoupling limit when we discuss the light neutralinos,
where there are 5 neutralinos with masses and compositions the same
as the nMSSM\footnote{The four decoupled neutralinos typically
consist of one heavy pair involving the $\tilde Z'$ and one linear
combination of $\tilde S_1$, $\tilde S_2$, $\tilde S_3$, as well as
two more states associated with the orthogonal combinations of
$\tilde S_{1,2,3}$ \cite{S-model,S-model_Higgs}. The latter can be
light, or even be the lightest neutralino in limiting cases. We do
not consider that possibility here.}.

\section{Direct Constraints}
\begin{figure}[t]
\resizebox{1.00\textwidth}{!}{
\includegraphics{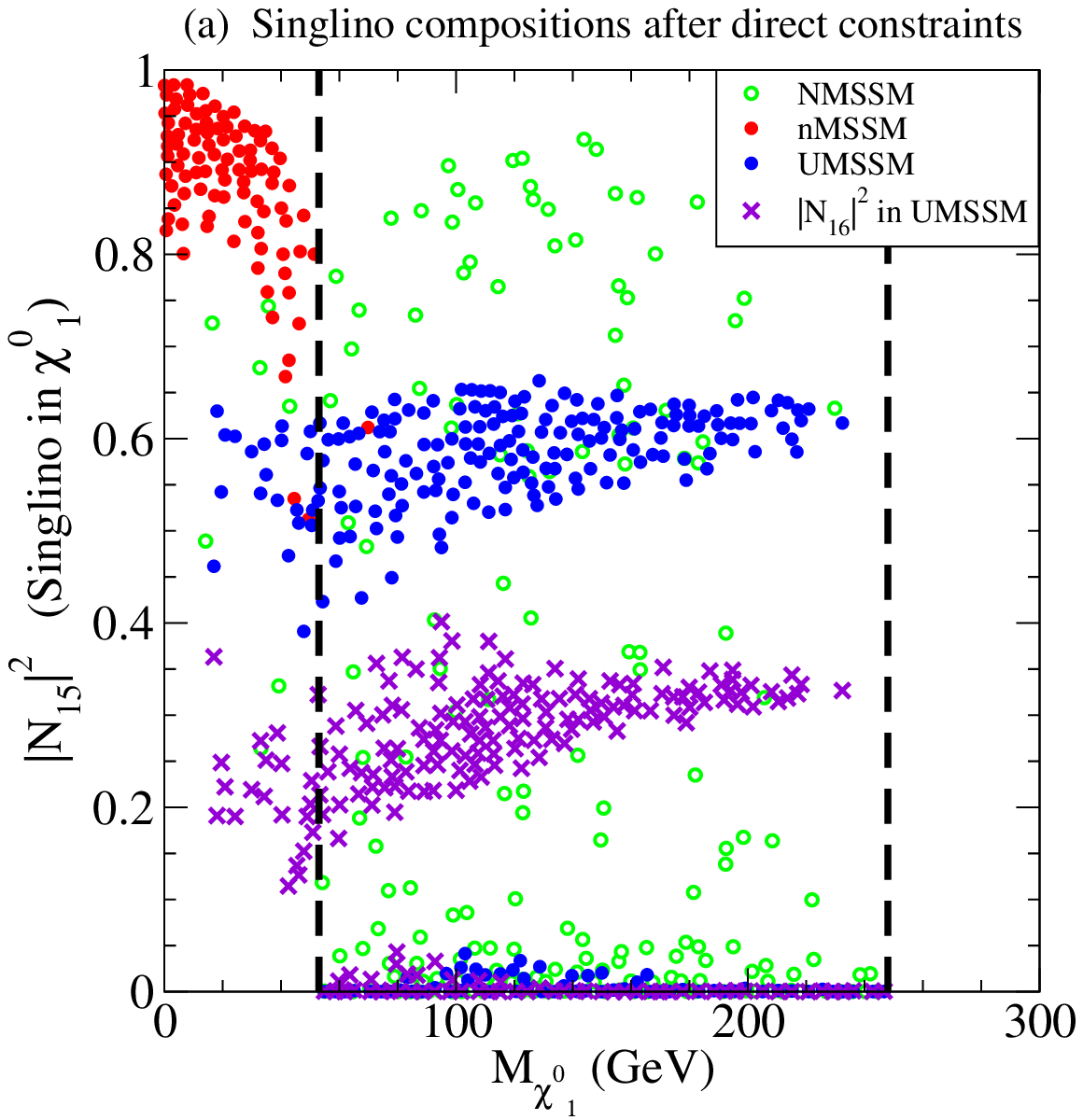}
\includegraphics{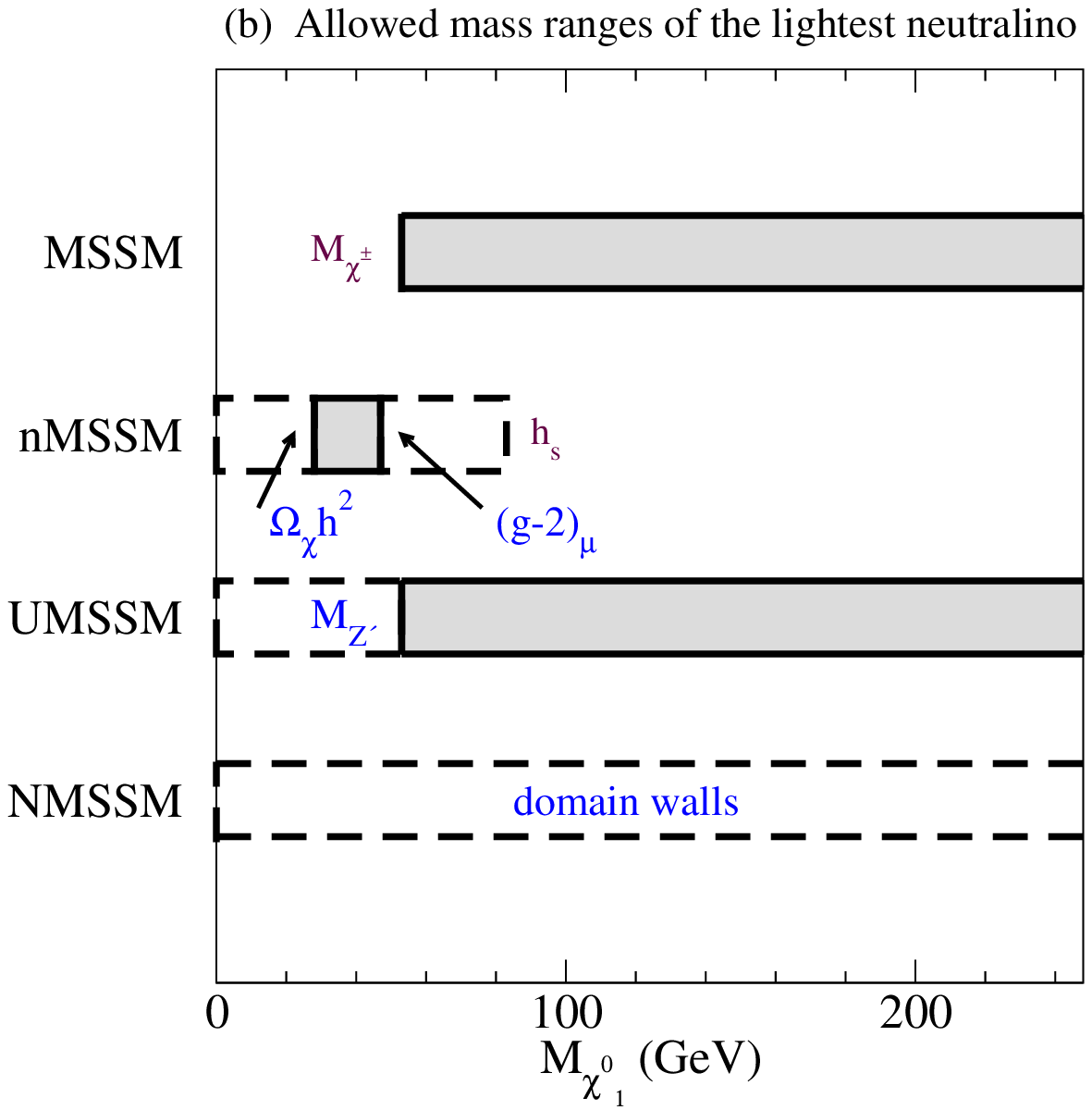}}
\caption{(a) Scatter plot of the $\chi^0_1$ singlino composition
($|N_{15}|^2$) versus $\chi^0_1$ mass ($M_{\chi_1^0}$) for various
models. The crosses represent the $Z'$-ino composition
($|N_{16}|^2$) for the UMSSM. The direct constraints are imposed.
The upper (lower) UMSSM singlino band corresponds approximately to
moderate (large) value of $s$, with the lightest neutralino being
MSSM-like for large $s$. (b) The allowed mass range of $\chi^0_1$
after applying direct constraints (model parameters, gaugino mass
unification $M_{1'} = M_1 \simeq 0.5 M_2$, $M_{\chi^\pm_1}$, $\Delta
\Gamma_Z$) and indirect constraints ($(g-2)_\mu$, $\Omega_{\chi^0_1}
h^2$, $M_{Z'}$, domain wall). The $M_{\chi^0_1}$ bounds in the nMSSM
are intended to be illustrative and are not necessarily
quantitatively precise.} \label{fig:tanb}
\end{figure}

The diagonalization of the neutralino mass matrix is accomplished
via a unitary matrix $N$ as \bea N^T M_{\chi^0} N = {\rm Diag}(
M_{\chi^0_1}, M_{\chi^0_2}, \cdots ). \eea The singlino ($\tilde S$)
composition of the lightest neutralino ($\chi^0_1$) is $|N_{15}|^2$.
Here we evaluate the bounds on the lightest neutralino mass
($M_{\chi^0_1}$) and the singlino component ($|N_{15}|^2$) in the
MSSM, the NMSSM, the nMSSM (also the decoupling limit of the
S-model), and the UMSSM with gaugino mass unification.

We require that the tree-level masses satisfy the direct LEP limits
of $M_{\chi^\pm_1} > 104$ GeV, and $\Gamma_{Z \to \chi_1^0 \chi_1^0}
< 2.3$ MeV ($95\%$ C.L.). Bounds from naturalness and perturbativity
constraints \cite{S-model,Miller:2003ay,neutralino_nMSSM} are also
imposed on the couplings of $0.1 \le h_s \le 0.75$ and $\sqrt{h_s^2
+ \kappa^2} \le 0.75$ (for the NMSSM)\footnote{An exact $h_s$ bound
(and its source) may be a little different depending on models, but
we assume a common bound for easy comparison. A lower bound on the
NMSSM $|\kappa|$ is fuzzy except that $\kappa \ne 0$ should be
satisfied to avoid an unacceptable Peccei-Quinn symmetry. We set
$|\kappa| \ge 0.1$ as a lower bound; results for a smaller
$|\kappa|$ can be described by an interpolation of the nMSSM result
($\kappa \to 0$ limit).}. The LEP2 Higgs mass bound of $m_h > 114$
GeV does not apply directly to the extended MSSM models where the
physical Higgses exist as mixtures of doublets and singlets
\cite{mixedhiggs}.

We choose a phase convention in which $\mu$ and the VEVs are real
and positive. $\kappa$ and the gaugino masses can in principle be
complex, but we restrict our considerations to real values. We scan
$M_2$, $\mu = 50 \sim 500$ GeV with a step size of $1$
GeV\footnote{We exclude very small $M_2$ or $\mu$ to avoid very
light non-singlino states. These are also excluded by the chargino
mass constraints.}, $s = 50 \sim 2000$ GeV with a step size of $5$
GeV, $\tan\beta = 0.5$, $1$, $1.5$, $2$, $10$, $50$, and $|\kappa| =
0.1 \sim 0.75$ with a step size of $0.05$ (for the NMSSM). The
gaugino mass unification relation $M_{1'} = M_1 = \frac{5}{3}
\frac{g_1^2}{g_2^2} M_2 \simeq 0.5 M_2$ is assumed.

The bounds that we obtain on the the lightest neutralino mass are
\bea 53 \mbox{ GeV} &\le& M_{\chi^0_1}~~ \le~~ 248 \mbox{ GeV  \quad
[MSSM]} \label{eqn:MSSMrange} \nn \\
0  \mbox{ GeV} &\le& M_{\chi^0_1}~~ \le~~ 248 \mbox{ GeV  \quad [NMSSM]} \nn \\
0  \mbox{ GeV} &\le& M_{\chi^0_1}~~ \le~~ {\tiny~} ~83 \mbox{ GeV
\quad [nMSSM, S-model]} \nn \\
0  \mbox{ GeV} &\le& M_{\chi^0_1}~~ \le~~ 248 \mbox{ GeV  \quad
[UMSSM]} \label{eqn:UMSSMrange} \nn \eea These results are only for
positive gaugino masses, but negative gaugino masses do not change
these ranges significantly\footnote{With $M_2 = -50 \sim -500$ GeV,
the $\chi^0_1$ mass ranges are $M_{\chi^0_1} = 39 \sim 254$ GeV
(MSSM), $M_{\chi^0_1} = 0 \sim 254$ GeV (NMSSM), $M_{\chi^0_1} = 0.4
\sim 96$ GeV (nMSSM), $M_{\chi^0_1} = 39 \sim 254$ GeV (UMSSM).}.
The mass bound dependence on $\tan\beta$ is quite sensitive in some
models. For example, at $\tan\beta \approx 1$, the NMSSM and the
UMSSM have massless $\chi^0_1$ while the nMSSM has an upper
$M_{\chi^0_1}$ bound \cite{main}; the MSSM violates the LEP2 $m_h$
constraint at this $\tan\beta$.
The condition for the massless (or very light) neutralinos can be
easily satisfied in extended MSSM models without the fine-tuning
which is required for the MSSM case \cite{main}.

Figure \ref{fig:tanb}(a) shows the $M_{\chi^0_1}$ dependence on the
singlino composition of $\chi^0_1$, $|N_{15}|^2$ (and also the
$Z'$-ino composition $|N_{16}|^2$ for the UMSSM). The dashed lines
are the MSSM bounds. Singlino dominance is typical in the $\chi^0_1$
of the extended MSSM models. Especially, when the $M_{\chi^0_1}$ is
much smaller than the MSSM lower limit\footnote{The relaxation
of the gaugino mass unification requirement would allow the MSSM also to
have a very light (bino-dominated) neutralino.
For the supernova constraint in this case, see Ref. \cite{Dreiner}.} ($M_{\chi^0_1} \sim 50$ GeV),
the singlino is always the dominant component.

\section{Discussion}
In the nMSSM (also in the S-model in the decoupling limit), because
of the small $\chi^0_1$ mass, the $Z$ pole is the dominant channel
\cite{neutralino_nMSSM, relic_U1}. To obtain an acceptable relic
density with only the $Z$ pole annihilation contribution, the lower
$M_{\chi^0_1}$ bound is $M_{\chi^0_1} \gsim 30$ GeV which is allowed
for only small $\tan\beta$, while the solution for a $2.4\sigma$
deviation from the SM of $(g-2)_\mu$ favors large $\tan\beta$, which allows
only small $M_{\chi^0_1}$. Nonetheless, a common solution was found
to exist in this model \cite{g-2_U1}. We refer to Ref.
\cite{lightpseudoNMSSM} for the interesting physics associated with
the light pseudoscalar Higgs boson including an alternative relic
density annihilation channel.

The NMSSM is disfavored by the non-observation of cosmological
domain walls predicted by the discrete symmetries of the model.
However, there is an approach to interpret the domain wall as the
dark energy \cite{domainwalldarkenergy}.

In the UMSSM, the mass of the new gauge boson $Z'$ is also an
important constraint. The Higgs singlet VEV $s$ should be large
enough to satisfy the experimental $M_{Z'}$ bound, although there
are some ways to get around it. (See Ref. \cite{main} for the
discussion.)

Approximate lightest neutralino mass ranges in the models considered
are illustrated in Figure \ref{fig:tanb}(b). The dashed regions are
disfavored by the indirect constraints ($(g-2)_\mu$,
$\Omega_{\chi^0_1} h^2$, $M_{Z'}$, domain wall). After the $M_{Z'}$
lower bound is imposed, the UMSSM bound becomes similar to the MSSM.
In the case of the nMSSM there is a tension between the $(g-2)_\mu$
constraint which favors small $M_{\chi^0_1}$ (or large $\tan\beta$)
and the relic density constraint which favors large $M_{\chi^0_1}$
(or small $\tan\beta$).

\section{Conclusion}
\label{conclusion}
Although Supersymmetry at the TeV scale is well-motivated, the MSSM
is just one of the possible realizations. In fact, the
$\mu$-problem suggests that the MSSM is incomplete. The solution to
the $\mu$-problem suggests that an appropriate direction to extend
the MSSM is to have an extra Higgs singlet whose VEV gives the
effective $\mu$-term of the EW scale. Extensions of the MSSM have
extra neutralinos, and the composition of the lightest neutralino
involves extra components beyond those of the MSSM. Because of this,
both the mass and couplings of the lightest neutralino are modified
from the MSSM. The lightest neutralino $(\chi^0_1)$ is interesting
both in particle physics (as the LSP) and cosmology (as the CDM),
and it is therefore important to study and compare properties of the
$\chi^0_1$ in extended MSSM models.

The properties of the CDM particle, even if it is the lightest
neutralino, may be quite different from the MSSM prediction. For
instance, it could be extremely light and/or dominated by the
singlino, which does not directly couple to SM particles except
Higgs doublets. Similar distinctions of models may occur in the
Higgs sector.
Even if a low energy Supersymmetry is correct, its realization may
depend on the model. The measurement of the mass of the lightest
neutralino and the determination of its couplings will be
particularly useful in testing the MSSM and its extensions at
colliders \cite{Hesselbach}.


\end{document}